# Atomically thin *p-n* junctions with van der Waals heterointerfaces

Chul-Ho Lee[1,2], Gwan-Hyoung Lee[3,4], Arend M. van der Zande[5], Wenchao Chen[6], Yilei Li[1], Minyong Han[7], Xu Cui[3], Ghidewon Arefe[3], Colin Nuckolls[2], Tony F. Heinz[1,8], Jing Guo[6], James Hone[3], and Philip Kim[1]*

[1]Department of Physics, Columbia University, New York, NY 10027, USA

[2]Department of Chemistry, Columbia University, New York, NY 10027, USA

[3]Department of Mechanical Engineering, Columbia University, New York, NY 10027, USA

[4]Department of Materials Science and Engineering, Yonsei University, Seoul 120-749, KOREA

[5]Energy Frontier Research Center (EFRC), Columbia University, New York, NY 10027, USA

[6]Department of Electrical and Computer Engineering, University of Florida, Gainesville, FL 32611-6130, USA

[7]Department of Applied Physics and Applied Mathematics, Columbia University, New York, NY 10027, USA

[8]Department of Electrical Engineering, Columbia University, New York, NY 10027, USA

*E-mail: pk2015@columbia.edu

**Semiconductor *p-n* junctions are essential building blocks for modern electronics and optoelectronics[1,2]. In conventional semiconductors, a *p-n* junction produces depletion regions of free charge carriers at equilibrium and built-in potentials associated with uncompensated dopant atoms. Carrier transport across the junction occurs by diffusion and drift processes defined by the spatial extent of this region. With the advent of atomically thin van der Waals (vdW) materials and their heterostructures, we are now able to realize a *p-n* junction at the ultimate quantum limit[3-10]. In particular, vdW junctions composed of *p*- and *n*-type semiconductors each just one unit cell thick are predicted to exhibit completely different charge transport characteristics than bulk junctions[10,11]. Here we report the electronic and optoelectronic characterization of atomically thin *p-n* heterojunctions fabricated using vdW assembly of transition metal dichalcogenides (TMDCs). Across the *p-n* interface, we observe gate-tuneable diode-like current rectification and photovoltaic response. We find that the tunnelling-assisted interlayer recombination of the majority carriers is responsible for the tunability of the electronic and optoelectronic processes. Sandwiching an atomic *p-n* junction between graphene layers enhances collection of the photoexcited carriers. The atomically scaled vdW *p-n* heterostructures presented here constitute the ultimate quantum limit for functional electronic and optoelectronic components.**

Heterostructures based on atomically thin vdW materials are fundamentally different from and more flexible than those made from conventional covalently-bonded materials: The lack of dangling bonds on the surfaces of vdW materials enables the creation of high-quality heterointerfaces without the constraint of atomically precise commensurability[9,12,13]. The ability to build artificial vdW heterostructures, combined with recent rediscoveries of TMDCs as atomically thin semiconductors[3-5], provides a route to a wide variety of semiconductor heterojunctions and superlattices. The strong light-matter interaction and the distinctive optical properties of TMDCs render them promising candidates for optoelectronic devices, such as photodiodes, photovoltaic cells, and light-emitting devices[3,8,14-19]. The availability of TMDCs with different band gaps and work functions allows for band gap engineering of heterostructures[20]. Furthermore, the ability to electrostatically tune the carrier densities and band alignments of two-dimensional (2D) semiconductors offers an alternative way to design functional heterostructures[21-24], and to understand the mechanisms underlying their operation. In this study, we have realized the ultimate limit of *p-n* junction scaling in a heterostructure consisting of semiconducting TMDC monolayers. We use individually contacted layers of *p*-type tungsten diselenide ($WSe_2$) and *n*-type molybdenum disulfide ($MoS_2$) to create an atomically thin *p-n* junction. The difference in work function and band gap between the two monolayers creates an atomically sharp, type II heterointerface[20].

Figure 1a shows a schematic of the vdW heterostructure of $MoS_2$ / $WSe_2$. Each TMDC monolayer consists of a hexagonally packed plane of transition metal atoms sandwiched between two planes of chalcogen atoms[6]. This heterojunction is created by vdW assembly of individual monolayers on a $SiO_2$ / Si substrate (see Sec. S1 in Supplementary Information (SI)). To measure the electrical properties of the junction, we fabricate metal contacts on each layer (optical image in Fig. 1a). Aluminium (Al) and palladium (Pd) are chosen to inject electrons and holes into the *n*-$MoS_2$ and *p*-$WSe_2$ layers, respectively[25]. A gate voltage ($V_g$) applied to the Si substrate adjusts the carrier densities in the each semiconductor layer. As shown in the inset of Fig. 1b, the individual $MoS_2$ and $WSe_2$ layers exhibit, respectively, *n*-type and *p*-type channel characteristics due to the unintentional doping present in each crystal. From the measured threshold voltages ($V_{th}$), we estimate (using *n* (electron) or *p* (hole) = $C_g |V_{th} - V_g|$, where $C_g = 1.23 \times 10^{-8}$ F·cm$^{-2}$) carrier densities of $1–3 \times 10^{12}$ cm$^{-2}$ at $V_g = 0$ V for both materials. Accordingly, a *p-n* junction is formed between the two atomically thin semiconductors. Figure 1b shows the current–voltage (*I–V*) curves of the junction at various gate voltages, as measured between metal electrodes on $WSe_2$ (D1) and $MoS_2$ (S1). We observe gate-voltage-tuned rectification of current as electrostatic doping modulates the density of free electrons and holes in the junction.

Although the observed *I–V* characteristics are similar to those of a conventional *p-n* junction diode, the underlying physical mechanism of rectification is expected to differ in view of the lack of a depletion region in the atomically thin junction. Figure 1c shows band profiles in the lateral (left) and the vertical (right) directions at a forward bias of 0.6 V, based on simulations for the electrostatic configuration of the device (see Sec. S2 in SI). We find that most of the voltage drop occurs across the vertical *p-n* junction, leaving no appreciable potential barriers in the lateral transport direction within each semiconductor layer. In contrast, under reverse bias, substantial potential barriers result from band bending in the lateral direction (see Fig. S3 in SI). Consequently, under forward bias, the current is governed by tunnelling-mediated interlayer recombination between majority carriers at the bottom (top) of the conduction (valence) band of $MoS_2$ ($WSe_2$). This process is prohibited in a typical bulk *p-n* junction because the depletion region spatially separates the majority carriers.

This unusual interlayer recombination can be described by two possible physical mechanisms or a combination of both: (i) Shockley-Read-Hall (SRH) recombination ($R \sim n_M p_W / \tau (n_M + p_W)$) mediated by inelastic tunnelling of majority carriers into trap states in the gap; and (ii) Langevin recombination ($R \sim B\, n_M\, p_W$) by Coulomb interaction. Here *R* is the recombination rate, $n_M$ and $p_W$ are, respectively, the electron and hole densities in $MoS_2$ and $WSe_2$, and $\tau$ and *B* are, respectively, the tunnelling-assisted recombination lifetime and the Langevin recombination constant[2] (see Sec. S2 in SI). The rectifying *I–V* characteristics are understood by increasing interlayer recombination rate at higher forward biases (see Fig. S4 in SI). In addition, the current under forward bias can be tuned by varying carrier densities through electrostatic gating; it is maximized with nearly balanced $n_M$ and $p_W$ at $V_g = 0$ (Fig. 1b), as predicted for both SRH and Langevin recombination. We estimate a lower bound for lifetime $\tau > 1$ μs from the measured current and the simulated carrier densities (see Sec. S3 in SI). The lifetime is relatively long because inelastic tunnelling processes occur between randomly stacked TMDC layers with lateral momentum mismatch[12]. This is beneficial for reducing the interlayer recombination in photocurrent generation, as discussed below.

Based on this understanding of the charge transport mechanism, we next explore the optoelectronic response of the atomically thin *p-n* junction. Figure 2a shows representative *I–V* curves and a colour plot of the photocurrent (inset) in the gate range of -30 – 30 V under white-light illumination. We observe a gate-tuneable photovoltaic response. In particular, the short-circuit current density becomes a maximum at $V_g$ = 0 V and decreases as varying the gate voltage in either polarity (the line profile at $V_{ds}$ = 0 V in the colour plot). The maximum photoresponsivity is ~2 mA/W, as measured using a focused 532-nm laser with a power of 0.8 μW (= 100 W/cm$^2$). Note that we observed a similar gate dependence for all light sources used, although the gate voltage was shifted by extrinsic substrate- or photo-induced doping effects[26].

To further elucidate the underlying physical processes, we performed spatial mapping of the photoresponse by scanning the light source in a confocal optical microscope. Figure 2b displays the resulting photocurrent map taken at $V_{ds}$ = 0 V. The strongest photoresponse is observed in the MoS$_2$ / WSe$_2$ junction area, indicating spontaneous charge separation occurring at the junction. This charge separation process is also compatible with the photoluminescence (PL) characteristics of the MoS$_2$ / WSe$_2$ junction. Figure 2c shows PL spectra obtained from the separated monolayers and from the overlapping layers forming the *p-n* junction. We find that emission from the direct gap transitions of both monolayer MoS$_2$ (1.88 eV) and monolayer WSe$_2$ (1.66 eV) is strongly quenched only at the MoS$_2$ / WSe$_2$ junction area. The integrated intensities decrease, respectively, by 81% and 98% compared to isolated MoS$_2$ and WSe$_2$ monolayers. The spatially resolved PL maps at the fixed band gap energies also reveal luminescence quenching in the *p-n* junction area (Fig. 2d). We attribute observed strong decrease in PL to the rapid separation of charge carriers in the junction region. While an energy transfer process could explain the decrease in PL by the material with the larger gap, the observed decrease of PL of both materials suggests a charge transfer mechanism. Note that we can exclude the photo-thermoelectric effect as the origin of the photocurrent: its contribution is minor compared to the measured photoresponse as a consequence of the small Seebeck coefficient and very slight temperature gradient across the vertical vdW interface (see Sec. S4 in SI)[27,28].

Spontaneous dissociation of a photogenerated exciton into free carriers can be driven by large band offsets for electrons ($\Delta E_c$) and holes ($\Delta E_v$) across the atomically sharp interface, as shown in the top panel of Fig. 2e. This charge separation may also be understood in terms of highly asymmetric charge transfer rates for electrons and holes in a heterostructure with type II band alignment: one process involves allowed transitions into the band, while the other process involves forbidden transitions into the gap. These charge transfer and separation processes are analogous to those considered in excitonic organic solar cells[11]. In atomically thin *p-n* junctions, the (allowed) charge transfer processes are expected to be fast and efficient since exciton (or minority carrier) diffusion is not required. Indeed, our observation of significant PL quenching and photocurrent generation in the *p-n* junction indicates that exciton dissociation at the junction is significantly faster than other non-radiative (or radiative) decay channels existing within the layers. For typical TMDC monolayers, such intralayer relaxation processes are present on the time scale of ~10 ps[3,29].

Unlike for the behavior of a conventional *p-n* junction after photoexcitation, for an atomically thin *p-n* junction, even after exciton dissociation, the majority carriers in each layer can undergo recombination though inelastic tunnelling. Such interlayer tunnelling-mediated recombination plays an important role in determining the photocurrent. The bottom panel of Fig. 2e represents interlayer SRH and Langevin recombination processes. Since the gate voltage tunes the majority carrier densities, we can model the gate modulation of the photocurrent observed in our experiments. Figure 2f displays plots of the simulated densities of majority and minority carriers in each layer as a function of gate voltages (top panel) under illumination. Then, the interlayer recombination rate would be proportional to $n_M\, p_W$ (middle panel in Fig. 2f) or $n_M\, p_W / (n_M + p_W)$ (bottom panel in Fig. 2f), depending on whether the dominant recombination mechanism follows Langevin or SRH behaviours, respectively.

These quantities, spatially averaged over the junction area, are minimized near $V_g = 0$ V. They increase with $|V_g|$ due to accumulation of one type of majority carrier. This indicates that both Langevin and SRH recombination provide reasonable qualitative descriptions of the sharply peaked photoresponse observed experimentally since the photocurrent is determined by difference of the recombination rate and the gate-independent generation rate. As shown in Fig. 2g, additionally, we obtain better quantitative agreement for the Langevin process (see Sec. S3 in SI). Note that SRH recombination cannot be ignored because of possible imperfections of the junction interface as well defects in the materials.

In our devices, the relatively low carrier mobility in the lateral transport channel results in a diffusion time for the majority carriers to leave the junction region as long as ~1 µs, a time scale comparable to the interlayer recombination lifetime. Since photocurrent collection competes with interlayer recombination, one strategy to reduce interlayer recombination losses is to increase the rate of collection of the photogenerated carriers. In order to realize this improvement, we employed graphene electrodes directly on the top and bottom of the vertical *p-n* junction (Fig. 3a). This allows carrier collection by direct vertical charge transfer, rather than through lateral diffusion. For comparison, we fabricated vdW stacks using various thicknesses of $MoS_2$ / $WSe_2$ layers in between two graphene electrodes (see Sec. S1 and S6 in SI)[30]. In this device geometry, graphene serves as a transparent vdW contact that also minimizes recombination as compared to typical metal contacts[8].

Figure 3b reveals that photocurrent (at $V_{ds} = 0$ V) is only observed in the region where the two active TMDC monolayers overlap with the top and bottom graphene electrodes (see also Sec. S7 in SI). Additionally, figure 3c shows *I–V* curves of the vertical *p-n* junction consisting of both monolayers in the dark and under 532-nm laser excitation. The device reveals linear *I–V* characteristics, apparently dominated by direct tunnelling between the two graphene electrodes (see Sec. S5 in SI). Nevertheless, we still observe a short-circuit current of ~70 nA at an excitation power of ~7 µW (= 920 W/cm$^2$), corresponding to a photoresponsivity of ~10 mA/W. This value for vertical charge collection is larger than that for a typical laterally-contacted device by a factor of ~5.

The *p-n* photodiode characteristics are improved significantly as the number of atomic layers in $MoS_2$ / $WSe_2$ heterojunctions increases and direct tunnelling between the top and bottom graphene electrodes is suppressed. Figure 3d shows the photoresponse characteristics of three devices with different junction thicknesses. As the thickness increases, the overall transport curve changes from a linear to a rectifying diode characteristic because direct tunnelling current decreases exponentially (see Sec. S5 in SI). In particular, the vertical *p-n* junction consisting of two multilayers (10 / 9 nm) shows a photovoltaic response with an open-circuit voltage of ~0.5 V (corresponding to ~26 mV/nm). Additionally, the photocurrent gradually increases with thickness up to ~120 mA/W because of increased light absorption. These photoresponse characteristics could also be tuned by varying the gate voltage (see Sec. S8 in SI).

We also investigate the external quantum efficiency (EQE) as a function of excitation energy for graphene-sandwiched *p-n* junctions with different thicknesses (Fig. 3e). The measured EQE at 532-nm is 2.4 %, 12 % and 34 % for monolayer, bilayer and multilayer *p-n* junctions, respectively. The quantity does not scale linearly with the junction thickness because of the thickness-dependent competition between generation, charge collection and recombination. Regardless of thickness, all three *p-n* junctions show the similar spectral response with two prominent peaks at 1.64 eV and 1.87 eV, corresponding to excitonic absorption edges of $WSe_2$ and $MoS_2$, respectively. This spectral photoresponse matches well with the absorption spectrum measured from the monolayer $MoS_2$ / $WSe_2$ stack (see Sec. S9 in SI), indicating that excitons generated in both semiconductor layers contribute to the photocurrent. These thickness- and wavelength-dependent results suggest a promising approach to realize efficient photovoltaic cells by multiple vdW assemblies of 2D semiconductors with different band gaps and work functions[6].

In conclusion, we have fabricated atomically thin *p-n* junctions from vdW-bonded semiconductor layers. Although the junction exhibits both rectifying electrical characteristics and photovoltaic response, the underlying microscopic processes differ strongly from those of conventional devices in which an extended depletion region play a crucial role. In particular, interlayer tunnelling recombination of the majority carriers across the vdW interface, which can be tuned by gating, is found to influence significantly both the electrical and optoelectronic properties of the junction. Further optimization of the band alignment, number of atomic layers, and interface lattice mismatching, will lead to unique material platforms for novel, high-performance electronic and optoelectronic devices.

**Methods Summary**

The $MoS_2$ / $WSe_2$ stacks were fabricated on 280-nm-thick $SiO_2$-coated Si substrates using a co-lamination and mechanical transfer technique. The thickness of TMDC layers was confirmed by atomic force microscopy, PL and/or Raman spectroscopy (see Sec. S6 in SI). The vertical stacks of graphene / $MoS_2$ / $WSe_2$ / graphene were prepared using a polymer-free vdW assembly technique with edge electrical contact to the graphene layers[30]. Further details about the fabrication processes are provided in Sec. S1 in SI. For model simulation of the device, Poisson and drift-diffusion equations are solved self-consistently using a 2D finite difference method (see Sec. S2 in SI). The photoresponse characteristics were investigated under white-light illumination and laser excitation. The quantitative analysis of the response was obtained with the focused laser excitation of known power. Additionally, scanning photocurrent measurements were performed using a confocal optical microscopy equipped with a two-axis scanning mirror or scanning mechanical stage. In these experiments, a 532-nm cw laser (Crystalaser) and a supercontinuum laser source (NKT photonics) was used as excitation light sources. The laser radiation was focused to a 0.7–1 μm FWHM spot through a 50× objective. The powers used for wavelength-dependent EQE measurements were in the range of 3–7 μW, where the devices exhibit linear response (see Sec. S10 in SI).

**Acknowledgements**

This work is supported as part of the Center for Re-defining Photovoltaic Efficiency Through Molecule Scale Control, an Energy Frontier Research Center funded by the U.S. Department of Energy, Office of Science, Office of Basic Energy Sciences under Award Number DE-SC0001085.

**Author contributions**

C.-H.L., G.-H.L., M.H., X.C. and G.A. fabricated and characterized the vdW *p-n* heterostructures. C.-H.L. and G.-H.L. performed device fabrication and transport measurements. A.M.v.d.Z. and C.-H.L. performed photocurrent measurements. Y.L. performed optical characterization. P.K., J.H., T.F.H. and C.N. advised on experiments. W.C. and J.G. provided theoretical support. C.-H.L. and P.K. wrote the manuscript in consultation with G.-H.L., A.M.v.d.Z., W.C., C.N., T.F.H., J.G. and J.H.

**Figure Captions**

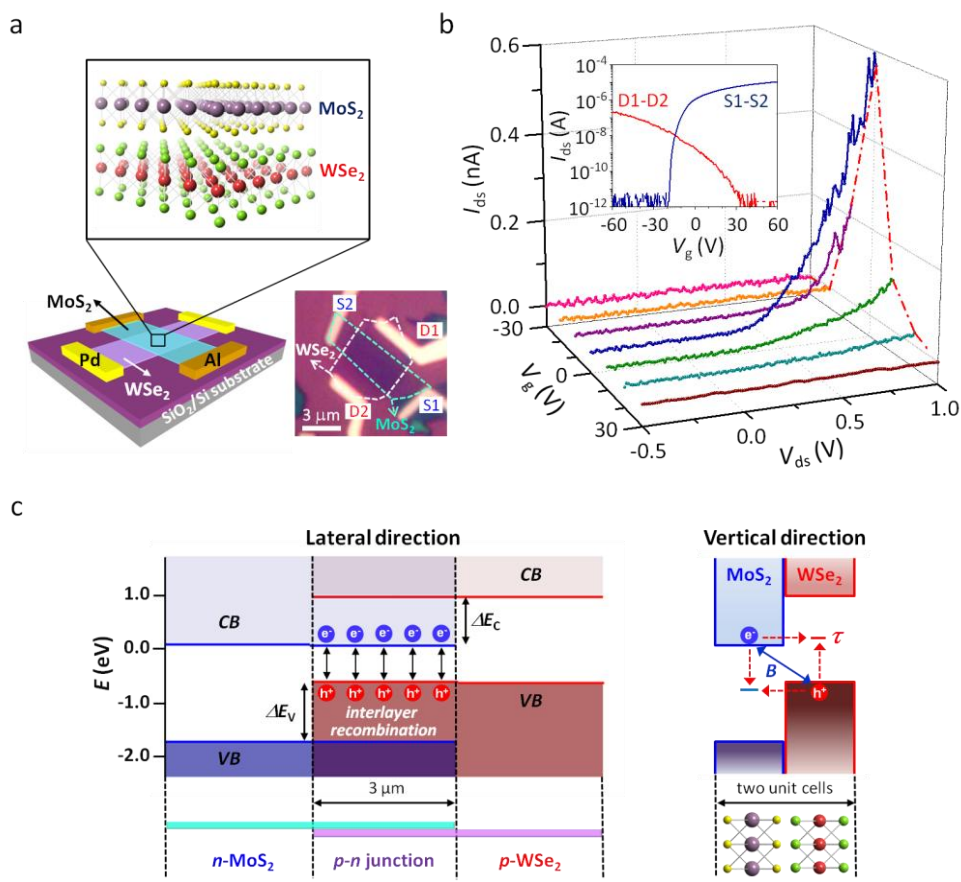

**Figure 1 | Charge transport in an atomically thin *p-n* heterojunction. a**. Schematic diagrams of a vdW-stacked MoS$_2$ / WSe$_2$ heterojunction and a laterally-contacted device. An optical image shows the fabricated device where D1 and D2 (S1 and S2) indicate the metal contacts for WSe$_2$ (MoS$_2$). **b**. *I–V* curves at various gate voltages measured across the junction (between contacts D1 (+) and S1 (−) in the optical image). The inset shows the gate-dependent transport characteristics at $V_{ds}$ = 0.5 V for individual monolayers of MoS$_2$ (S1-S2) and WSe$_2$ (D1-D2). **c**. Band profiles in the lateral (left panel) and the vertical (right panel) direction, obtained from electrostatic simulations. Under forward bias ($V_{ds}$ = 0.6 V in this figure), electrons in conduction band (CB) of MoS$_2$ and holes in valence band (VB) of WSe$_2$ undergo interlayer recombination via SRH (red dashed arrows) or Langevin (blue arrow) mechanism, contributing the dark current. Note that there is no significant band bending in the lateral transport direction. Band offsets for electrons ($\Delta E_c$) and holes ($\Delta E_v$) across the junction are indicated in the left band profile.

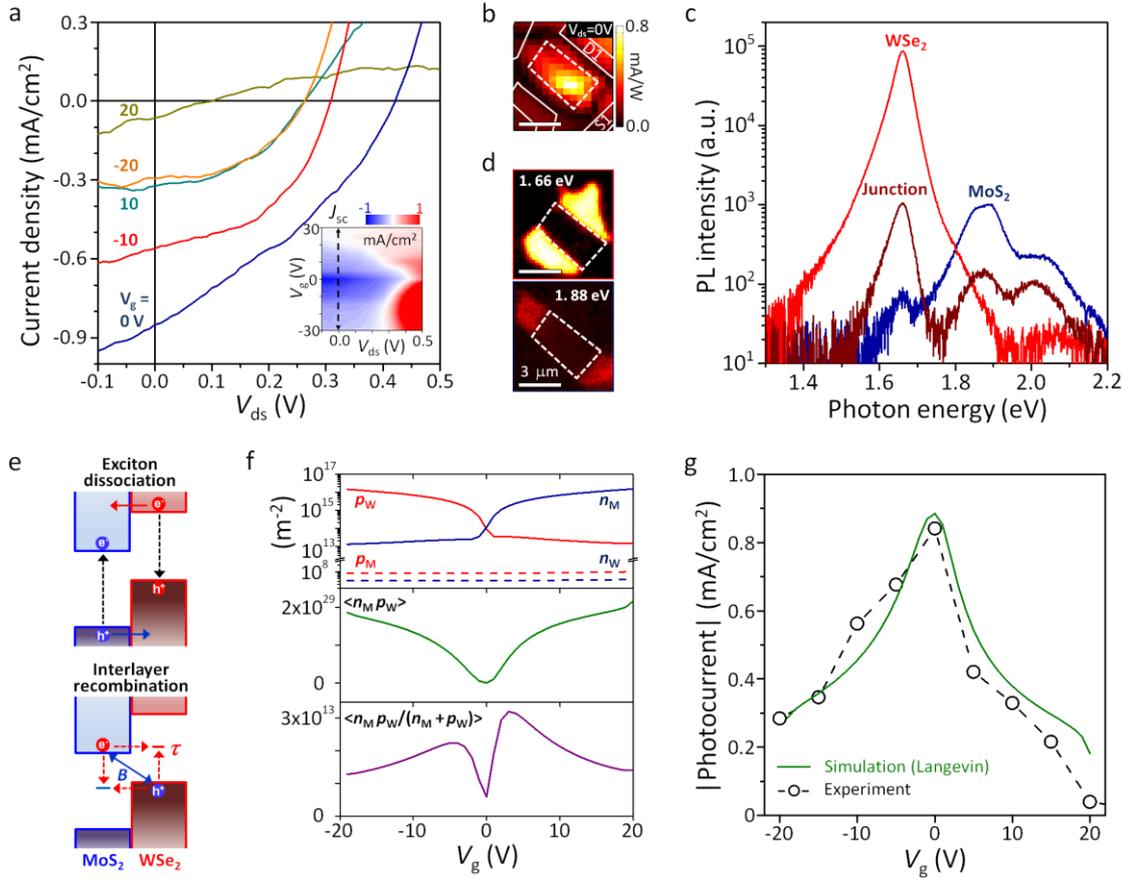

**Figure 2 | Gate-tuneable photovoltaic response. a**. Photoresponse characteristics at various gate voltages under white light illumination. The inset shows a colour plot of the photocurrent as a function of voltages $V_{ds}$ and $V_g$, where blue and red colours represent negative and positive photocurrent, respectively. The dashed line represents the profile of short-circuit current density ($J_{sc}$) at $V_{ds} = 0$ V. **b.** Photocurrent map of the device presented in Fig. 1**a** for $V_{ds} = 0$ V and 532-nm laser excitation. The junction area and metal electrodes are indicated by the dashed and solid lines, respectively. **c**. PL spectra measured from the isolated monolayers and the stacked junction region. **d.** PL spatial maps for emission at 1.66 eV and 1.88 eV, corresponding to direct gap transitions of monolayer $WSe_2$ and $MoS_2$, respectively. The junction area is indicated by the dashed lines. **e.** Schematic illustrations of exciton dissociation (top panel) and interlayer recombination processes (bottom panel). In the top panel, the horizontal and vertical arrows represent charge transfer and intralayer recombination processes, respectively. In the bottom panel, the red and blue arrows indicate SRH and Langevin recombination processes, respectively. **f.** Simulations of the gate-voltage-dependent majority ($p_W$ for holes in $WSe_2$ and $n_M$ for electrons in $MoS_2$) and minority carrier ($p_M$ for holes in $MoS_2$ and $n_W$ for electrons in $WSe_2$) densities in each layer (top), spatially averaged $<n_M p_W>$ (middle) and $<n_M p_W/(n_M + p_W)>$ (bottom). **g.** The measured and simulated photocurrent (at $V_{ds} = 0$ V) as a function of gate voltages. The experimental data show good quantitative agreement with the simulated curve from Langevin recombination mechanism (for $B = \sim 2 \times 10^{-10}$ m$^2$/s).

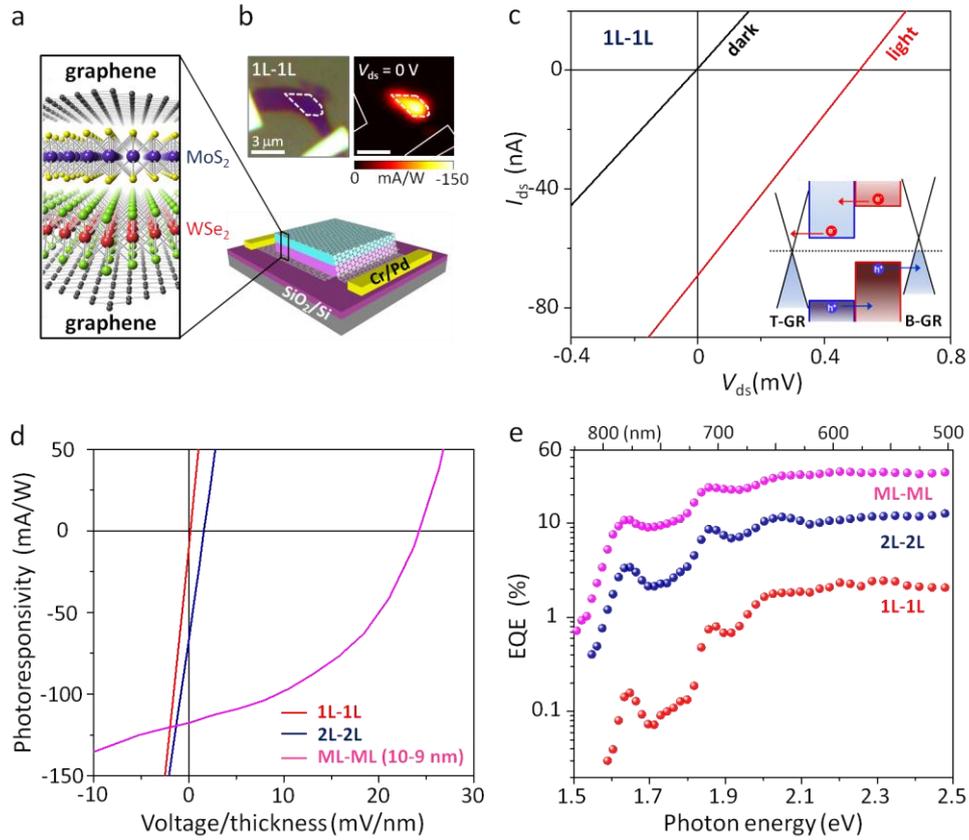

**Figure 3 | Graphene-sandwiched vdW *p-n* heterojunctions. a**. Schematics of a $MoS_2$ / $WSe_2$ junction sandwiched between the top and bottom graphene electrodes. **b**. Optical image and the corresponding photocurrent map (for $V_{ds}$ = 0 V) of the graphene-sandwiched monolayer *p-n* junction device (1L-1L). Photocurrent is uniformly observed only in the junction area indicated by the dashed lines, separated from metal electrodes indicated by the solid lines. **c**. *I–V* curves of the device in **b**, measured in the dark and under 532-nm laser excitation. The inset shows a schematic band structure where exciton dissociation and charge collection processes are indicated by the horizontal arrows. The vertical arrows represent the possible intralayer recombination processes. Note that the bottom graphene (B-GR) is slightly *p*-doped from the $SiO_2$ substrate. **d**. Photoresponse characteristics of graphene-sandwiched *p-n* junctions with different thicknesses. For comparison, the applied voltages (horizontal axis) are normalized with respect to the junction thicknesses. **e**. EQE plots as a function of excitation energy (wavelength) for the devices in **d**. The laser powers were used in the range of 3–7 μW, corresponding to 380–890 W/cm$^2$. Results are shown for devices composed of monolayer/monolayer (1L-1L), bilayer/bilayer (2L-2L) and multilayer/multilayer (ML-ML (10-9 nm)) junctions.

# *Supplementary Information*

## S1. Experimental methods

We utilized two similar transfer procedures to fabricate van der Waals vertical *p-n* heterostructures.

### Co-lamination & mechanical transfer technique for a vertical stack of $MoS_2$ / $WSe_2$

For the laterally metal-contacted *p-n* junction device (Fig. 1 and Fig. 2), we utilized a co-lamination and mechanical transfer technique to fabricate the vertically stacked $MoS_2$ / $WSe_2$ heterostructure[1,2]. First, we used mechanical exfoliation with scotch tape to obtain monolayer $WSe_2$ on a 280-nm-thick $SiO_2$-coated Si substrate. Separately, we used mechanical exfoliation to obtain monolayer $MoS_2$ on a Si substrate coated with a polymer bilayer consisting of thin water-soluble polyvinyl acetate (PVA) and 280-nm-thick poly(methylmethacrylate) (PMMA). We released the PMMA film with $MoS_2$ flakes from the substrate by dissolving the PVA in water and transferred the film onto a polydimethyl siloxane (PDMS) stamp. Finally, we used a micromanipulator to place the $MoS_2$ layer on top of the $WSe_2$ monolayer through an aligned transfer procedure[1]. This transfer process was performed at a substrate temperature of 120 °C. After dissolving the PMMA in Acetone, we fabricated electrical contacts using e-beam lithography. Importantly, we fabricated separate contacts for the individual layers to minimize the Schottky barriers. Specifically, the $MoS_2$ was contacted with Al / Cr / Au (40 / 1 / 50 nm) while the $WSe_2$ was contacted with Pd / Au (20 / 30 nm).

### vdW assembly of a graphene-sandwiched $MoS_2$ / $WSe_2$ heterostructure

For the graphene-sandwiched *p-n* junction device (Fig. 3), we utilized a recently published transfer technique utilizing the van der Waals adhesion of layers to "pickup" other layered materials while avoiding use of polymers and solvents[3]. We started by exfoliating h-BN layers with thickness of 10−30 nm on the ~1-μm-thick poly-propylene carbonate (PPC) film on silicon, and separately exfoliating graphene and TMDC monolayers onto the $SiO_2$ (280 nm) / Si substrates. We mechanically removed the PPC film with h-BN flakes from the Si substrate and placed the film onto a transparent elastomeric stamp (PDMS) fixed on a glass slide. To pick up the graphene layer (T-GR), we attached the inverted PDMS stamp to a micromanipulator and aligned the h-BN flake over a chosen exfoliated layer, brought the two flakes into contact, and then picked up the target graphene flake via a vdW interaction. This pick-up process was performed at a substrate temperature of 40 °C. The same process was repeated for the $MoS_2$ and $WSe_2$ layers to build the vertical *p-n* junction. Finally, we transferred the whole stack of h-BN / graphene / $MoS_2$ / $WSe_2$ on the PPC / PDMS stamp onto another graphene flake (B-GR) on the $SiO_2$ / Si substrate by melting the PPC film at 90 °C, resulting in the vdW stack of h-BN / graphene / $MoS_2$ / $WSe_2$ / graphene on the $SiO_2$ / Si substrate.

We used e-beam lithography to etch and add metal contacts to the heterostructures. First, we used PMMA/hydrogen silsesquioxane (HSQ) bilayer as an etch mask to define the junction in the transferred multi-stack. We performed a dry etch using an inductively coupled plasma etcher (Oxford ICP 80 system) with a mixture of $O_2$ and $CHF_3$ reactive gases. The typical etch rate was ~30 nm/min. After this process, we liftoff the PMMA / HSQ etch mask in acetone. We fabricated separate edge contacts by evaporating Cr / Pd / Au (10 / 15 / 50 nm) electrode to contact the top and bottom graphene layers (See Fig. S1)[3].

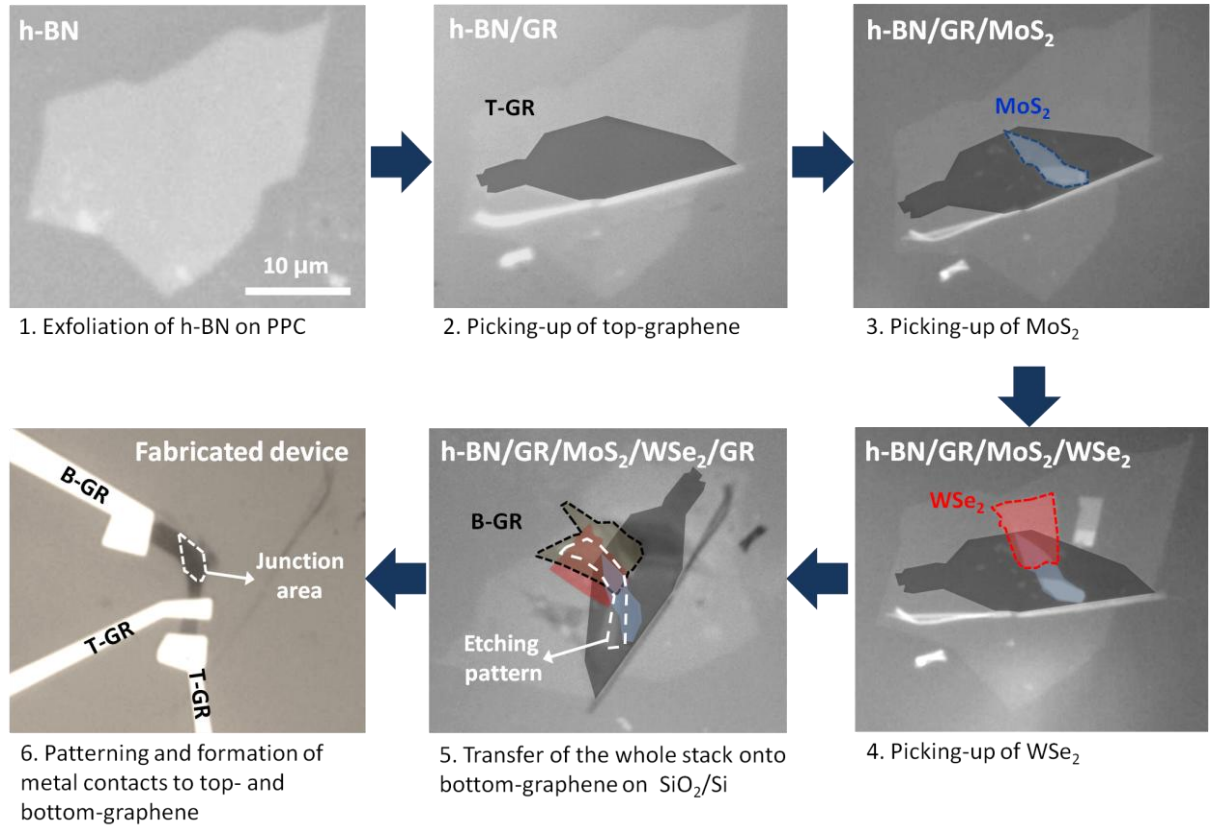

**Figure S1 | Fabrication of a graphene-sandwiched *p-n* heterojunction device.** A series of optical images, representing each fabrication step for the graphene-sandwiched $MoS_2$ / $WSe_2$ monolayer device described in the main text (Fig. 3). In each optical image, the boundaries of individual atomic layers are indicated by the dashed lines.

**Electrical and photocurrent measurements**

We performed electrical measurements by utilizing Keithley 2400 power supplies to applying drain and gate voltages and a current amplifier (DL 1211) for current measurements.

The photocurrent measurements were carried out using three different methods. In Method 1, we exposed the sample to wide-field and broad spectrum light from a tungsten lamp through a microscope objective (10×). The data obtained in Fig. 2a utilized Method 1. Methods 2 and 3 utilized scanned photocurrent measurements using either a 532-nm cw laser (Crystalaser) or a supercontinuum source (NKT photonics) with a monochromer (7 nm FWHM). In both methods, the laser was focused down onto the sample through a objective (50×) to a ~ 0.7−1 μm FWHM excitation spot. In Method 2, we performed scanning measurements using a scanned mechanical stage, while in Method 3, the we performed scanning measurements using a confocal scanning laser setup. The data in Fig. 2b were obtained by Method 2, while the data in Fig. 3b were obtained by Method 3.

## S2. Device model simulation

Figure S2 shows a schematic the simulated device structure where thicknesses of the gate oxide and metal electrodes are 280 nm and 200 nm, respectively. The spacing between source and drain electrodes is 9 μm and the length of the MoS$_2$ / WSe$_2$ junction is $L_j = 3$ μm, closely following the device dimension in Fig. 1 and Fig. 2. The interlayer spacing between WSe$_2$ and MoS$_2$ is 7 Å, assuming the vdW distance between the two layers. The width of the device (in the out-of-plane direction) is $W = S_j / L_j \approx 2.93$ μm, where $S_j \approx 8.8$ μm$^2$ is the junction area of the experimental device. Because the device is wide, a self-consistent two-dimensional (2D) simulation is used as described below. A scaling factor of $1 / S_F$ is used in the lateral transport direction, as discussed and validated later, for computational efficiency.

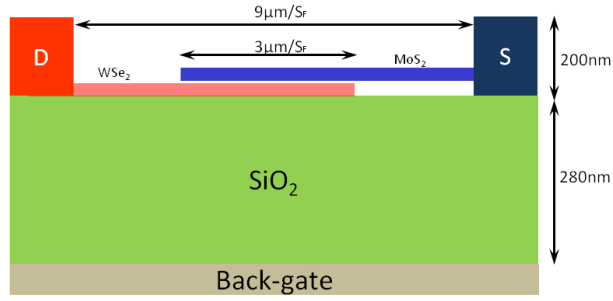

**Figure S2** | Schematic of the simulated device structure. The physical dimensions used for model simulation are indicated. .

Poisson and drift-diffusion equation are solved self-consistently by a 2D finite difference method to investigate device operation mechanisms. Poisson equation is listed as follows,

$$\nabla \cdot (\varepsilon_r \nabla V) = \frac{q}{\varepsilon_0}(n - p + N_A - N_D) \qquad (1)$$

where $\varepsilon_r$ and $\varepsilon_0$ are, respectively, the relative dielectric constant and vacuum permittivity. $V$ is the voltage, $q$ is the elementary charge. And, $n$ and $p$ are the electron density and hole density, respectively. $N_A$ is $p$-type doping concentration in WSe$_2$ and $N_D$ is $n$-type doping concentration in MoS$_2$. The drift-diffusion equation and current continuity equation are shown as follows

$$J_n = -qn\mu_n \nabla V + qD_n \nabla n \qquad (2)$$

$$J_p = -qp\mu_p \nabla V - qD_p \nabla p \qquad (3)$$

$$\frac{1}{q}\nabla \cdot J_n - R_n + G_n = 0 \qquad (4)$$

$$-\frac{1}{q}\nabla \cdot J_p - R_p + G_p = 0 \qquad (5)$$

where $J_n, \mu_n, D_n$ are the electron current density, the electron mobility and the electron diffusion coefficient, respectively. $J_p, \mu_p, D_p$ are the corresponding values for holes. $G_n$ and $G_p$ are electron and hole generation rates by optical illumination. $G_n$ and $G_p$ become zero in the dark. $R_n$ and $R_p$ are electron and hole recombination rates, which plays an important role in determining the photocurrent as well as the dark current.

Considering the presence of trap states and low photoluminescence (PL) quantum yield of semiconducting TMDCs[4], intralayer recombination can be dominated by the Shockley-Real-Hall (SRH) recombination mechanism[5,6],

$$R_{intra} = \frac{np - n_0 p_0}{\tau_{intra,p}(p+p_1) + \tau_{intra,n}(n+n_1)} \sim \frac{np}{\tau_{intra,}(n+p)} \quad (6)$$

where $n_0$ and $p_0$ are electron and hole density at equilibrium condition, $n_1$ and $p_1$ are electron and hole density for the case in which the Fermi level falls at $E_t$, and $\tau_{intra}$ is intralayer carrier lifetime.

Meanwhile, interlayer recombination can be described by trap-assisted SRH or Coulomb-interacting Langevin processes[7], equations for two recombination mechanisms are expressed as follows,

$$R_{inter,SRH} \sim \frac{n_M p_W}{\tau_{inter}(n_M + p_W)} \quad (7)$$

$$R_{inter,L} \sim B n_M p_W \quad (8)$$

where $\tau_{intra}$ is interlayer tunneling life time and $B$ is Langevin recombination constant. Note that SRH recombination rate is simplified because $n_0, p_0, n_1$ and $p_1$ are typically orders of magnitude lower than the values of $n$ and $p$. Also, the difference between $\tau_n$ and $\tau_p$ is neglected for simplicity, and differentiating these two values does not change the qualitative conclusions.

It should be noted that drift-diffusion equations and current continuity equations are solved in two individual monolayers (lateral direction) and their overlapped region (vertical direction). In the lateral direction, we use mobility measured in individual monolayer devices before the junction is formed. In the vertical direction, although interlayer transport is dominated by quantum tunnelling, we used a phenomenological interlayer carrier mobility $\mu_{inter}$ to describe the interlayer carrier transport between $MoS_2$ and $WSe_2$ monolayers,

$$\mu_{inter} = \frac{v}{E} = \frac{a/\tau_t}{V/a} \quad (9)$$

where $a$ is spacing between $MoS_2$ and $WSe_2$, $V$ is voltage drop between two monolayer materials, and $\tau_t$ is the tunneling time between two layers. The value of $V$ can be approximately set as the difference between the conduction (valence) band edge of $MoS_2$ and valence (conduction) band of $WSe_2$ divided by elementary charge for electron (hole) transport. Numerical calculation shows that the value varies about 10 % in the bias range of interest. Note that phenomenological mobility used in our simulation results in faster interlayer charge transfer rate than intralayer recombination rate as expected from experimental observation of strong PL quenching in the $MoS_2$ / $WSe_2$ heterojunction.

Because the large aspect ratio between the lateral size and the vertical interlayer distance, numerical challenges need to be taken care of in the simulation. We developed and validated a scaling method, which scales down the dimension in the transport direction by a scaling factor of $1 / S_F$ to reduce computational cost. The intralayer mobility in the horizontal direction is scaled down by factor $1 / S_F$ to maintain same parasitic resistance. The recombination and generation rates are scaled up by factor $S_F$ to maintain same current per unit width. The phenomenological interlayer mobility is scaled up by a factor of $S_F$. We have numerically tested that the physical quantities of interest, including the dark and photoresponse $I$–$V$ characteristics of the junction and the charge density in the junction region, are insensitive to the scaling factor if $S_F \leq 30$. The electrostatic screening length in the horizontal direction is considerably shorter than 100 nm due to atomic thickness of $WSe_2$ and $MoS_2$. To further speed up simulation, a non-uniform numerical grid is used in the vertical direction for the numerical solution of Poisson equation.

## S3. Electrical characteristics of laterally metal-contacted *p-n* heterojunction in the dark

Figure S3 presents band diagrams in the lateral direction under reverse ($V_{ds} = -0.6$ V), zero (0 V), forward (0.6 V) biases at $V_g = 0$ V, respectively. At a reverse bias, holes in *p*-doped $WSe_2$ and electrons in *n*-doped $MoS_2$ are depleted, and the quasi-Fermi level of $MoS_2$ in the non-overlapped region is close to the conduction band edge, and that of $WSe_2$ is close to the valence band edge. Depletion in the junction region and quasi-Fermi level splitting at the reverse bias results in large band bending between overlapped and non-overlapped regions in the lateral direction. On the other hand, at a forward bias, both holes in *p*-doped $WSe_2$ and electrons in *n*-doped $MoS_2$ are accumulated. Therefore, band bending in the lateral direction is negligibly small as shown in the bottom panel of Fig. S3 as well as in Fig. 1c.

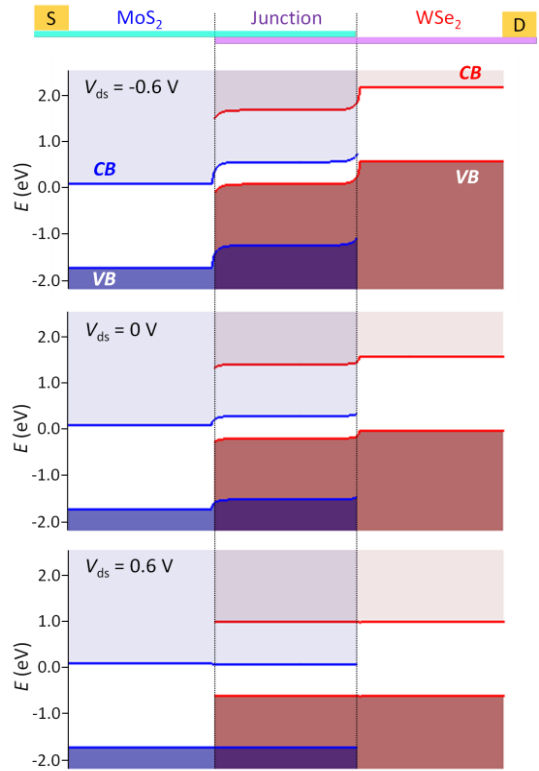

**Figure S3** | Band diagrams in the lateral transport direction at different biases, obtained from model simulation considering electrostatics. (top panel: $V_{ds} = -0.6$ V, middle panel: 0 V, bottom panel: 0.6 V)

In a conventional bulk *p-n* junction, the forward bias current is determined by recombination followed by diffusion of majority carriers over electrical potential barriers across the depletion region, resulting in an exponential *I–V* characteristic. In contrast, due to lack of a depletion region, charge transport in an atomically thin junction is governed by interlayer recombination processes between two majority carriers accumulated in each layer, and thus *I–V* curves do not show an exponential characteristic even for an ideal junction. In our model simulation, we examined two recombination mechanisms described above. Figure S4 shows comparison of experimental and theoretical dark *I–V* curves at $V_g = 0$ V. Both SRH and Langevin mechanisms show good agreements with the experimental data. For quantitative matching, we estimated the interlayer tunneling life time ($\tau_t$) of 40 μs and Langevin recombination constant (*B*) of $1.67 \times 10^{-13}$ m$^2$/s.

Unlike an ideal case in simulation, in real devices, parasitic resistances are likely to exist in both metal (Pd)-WSe$_2$ and metal (Al)-MoS$_2$ contacts, as well as lateral charge transport in WSe$_2$ and MoS$_2$ monolayers. We indeed found that the simulated *I–V* characteristics are sensitive to the parasitic resistance, which can reduce the splitting of quasi-Fermi-energy levels of majority carriers in the overlapped junction region and thereby results in lower majority carrier densities. For example, current decreases especially at large biases when we include the Schottky barrier of 0.3 eV at the Pd-WSe$_2$ junction. Consequently, we estimated smaller tunnelling lifetime (~microsecond time scale (> 1 μs)) and larger Langevin recombination constant (~$2.0 \times 10^{-10}$ m$^2$/s), in order to obtain the forward bias current measured in our device. Although it is very difficult to get a quantitative number precisely, it should be noted that the microsecond time sale obtained here is much longer than those of any other competing processes including intralayer radiative (or nonradiative) recombination rate and charge transfer rate (for exciton dissociation) as discussed in the main text[8].

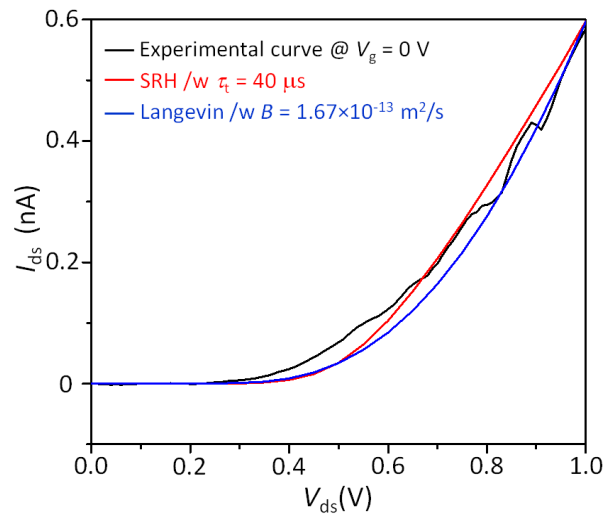

**Figure S4 |** Comparison of the measured and the simulated *I–V* curves in the dark at $V_g = 0$ V. In our model simulation, both SRH and Langevin mechanisms are considered, showing good agreements with the measured *I–V* curve.

This interlayer tunneling recombination plays also a dominant role in determining the photocurrent as discussed in the main text (Fig. 2f and Fig. 2g). As shown in Fig. 2f, for both SRH and Langevin mechanisms, the recombination rate is minimized near $V_g = 0$ V, resulting in the maximum photocurrent under a constant exciton generation rate. The recombination rate increases as varying the gate voltage to either polarity due to accumulation of one type of majority carrier. Although both mechanisms are reasonable to qualitatively explain the experimental results, the Langevin type shows excellent quantitative agreement with the recombination constant ($B$) of ~$2.0 \times 10^{-10}$ m$^2$/s, as shown in Fig. 2f.

## S4. Origin of photocurrent: Photovoltaic *vs.* Photo-thermoelectric effects

We compute the magnitude of the photo-thermoelectric effect to compare the relative contribution to the measured photocurrent in our *p-n* heterojunction devices[9,10]. The photovoltage generated by the thermoelectric effect can be formulated as

$$V_{PTE} = (S_{MoS2} - S_{WSe2})\Delta T \qquad (10)$$

where $S_{MoS2}$ and $S_{WSe2}$ are Seebeck coefficients (thermoelectric power) of MoS$_2$ and WSe$_2$ layers, respectively, and $\Delta T$ is the temperature difference between the two layers. From the Mott formula[11], we can write $S$ as

$$S = -\frac{\pi^2 k_B^2 T}{3e} \frac{1}{G} \frac{dG}{dE}\bigg|_{E=E_f} \qquad (11)$$

where $T$ is the sample temperature, $k_B$ is the Boltzmann constant, and $E_f$ is the Fermi energy. We calculate the Seebeck coefficient using the conductance data of the individual monolayers from the inset to Fig. 1b, and rewriting the equation[9] as

$$\frac{1}{G}\frac{dG}{dE} = \frac{1}{G}\frac{dG}{dV_g}\frac{dV_g}{dE} \qquad (12)$$

where the first two terms ($1/G$ and $dG/dV_g$) come from the conductance data, and the Fermi energy can be expressed[12] in terms of the gate voltage using

$$E_f = \frac{\hbar^2 \pi n}{2m_e} \qquad (13)$$

$$n = \frac{1}{e}C_g(V_g - V_0) \qquad (14)$$

where $n$ is the density of electrons/holes in the material, $m_e = 0.35m_0 \times m_h = 0.428m_0$ is the electron/hole mass at the $k$ point (using the numbers for MoS$_2$ only), and $C_g = 1.23 \times 10^{-4}$ F·m$^{-2}$ is the capacitance per unit area of a back-gated 2D material on a 280-nm-thick SiO$_2$ / Si substrate. With these numbers, we compute the Seebeck coefficient of two TMDC monolayers at $V_g = 0$ V as $S_{MoS2}$ ~ -3.3 mV/K and $S_{WSe2}$ ~ 4.5 mV/K. Then, total contribution becomes to be $\Delta S$ ~ -7.8 mV/K.

Since we do not have a good measure for thermal conductivity across a disparate vdW interface, it is difficult to compute the temperature difference in our samples. However, at the ~ 1–5 μW powers used in our measurements, it is very unlikely to be above 1 K[9,10,13,14]. If we compare the value of the photothermal voltage (< 10 mV) to the measured open-circuit voltage of 500–700 mV, it is clear that the photothermal contribution is small compared to the photovoltaic effect. This calculation was performed for the vertical MoS$_2$ / WSe$_2$ junction, however the Seebeck coefficients will not change drastically for the laterally-contacted heterojunction, so the contribution should be small in all geometries studied here.

## S5. Electrical characteristics of graphene-sandwiched *p-n* junctions with different thicknesses

The photoresponse *I–V* characteristics of graphene-sandwiched *p-n* junctions are modelled by a simple circuit model as shown in Fig. S5a. $R_s$ and $R_t$ are series resistance and the tunnelling resistance between the top and bottom graphene contacts, respectively. $I_0$ is current source related to photogenerated current. Despite simplicity of the model, it explains the qualitative features of the *I–V* characteristics depending on the number of layers as shown in Fig. S5b.

The shunt resistance is due to tunneling between two graphene contacts separated by the atomically thin *p-n* junction. Since tunnelling current is exponentially dependent on tunnelling distance, the shunt resistance exponentially depends on the number of layers as shown below,

$$R_t = R_0 e^N \qquad (10)$$

where *N* is number of layers. By carefully choosing parameters, we can get the simulation results with a quantitative agreement to experimental measurements. The parameters used are $R_0 I_N = R_s I_N = 5 \times 10^{-4} (V)$ and $I_0 / I_N = 10$, where $I_N$ is a normalization current. The diode shows the *I–V* characteristic same as the multilayer *p-n* junction in the dark, but with a normalization unit that results in *I* (at *V* = 0.7 V) / $I_N \approx 80$. Piecewise function is used to fit the dark *I–V* characteristics. The simple model captures the qualitative features of the nearly linear *I–V* characteristics of the monolayer junction, the increase of the open-circuit voltage and the transition to the diode-like *I–V* with increasing the number of layers.

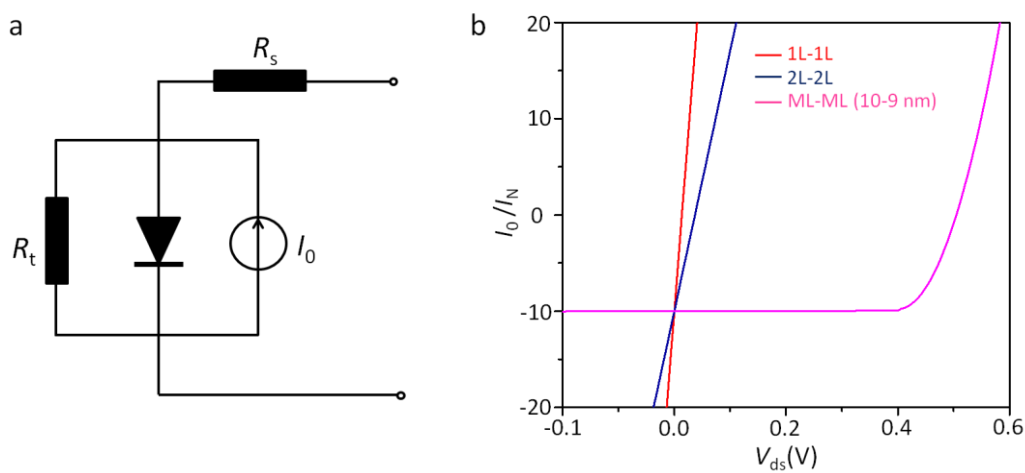

**Figure S5 | a.** Circuit diagram for modelling the graphene-sandwiched *p-n* heterojunction device, where the series resistance ($R_s$) and shunt tunneling resistance ($R_t$) are included. **b.** Simulated *I–V* curves of the graphene-sandwiched *p-n* heterojunctions with different thicknesses described in the main text (Fig. 3d).

## S6. Identification of the number of TMDC layers

Figure S6a and S6b show PL spectra of $MoS_2$ and $WSe_2$ with monolayer and bilayer thicknesses, respectively, from which one can identify the number of layers. The monolayer exhibits strong PL emission corresponding to the direct gap transition at ~1.88 (1.66) eV for $MoS_2$ ($WSe_2$). For the bilayer cases, due to indirect gap transition, the direct gap emissions are significantly reduced, and the emissions at lower energies are slightly appeared[4]. In addition to PL, Raman spectra can be used to determine the number of layers. For the $MoS_2$ layers, as shown in Fig. S6c, we observe strong signal from both the in-plane ($E^1_{2g}$) and the out-of-plane ($A_{1g}$) vibration, clearly showing the shift of Raman frequencies with increasing the layers[15].

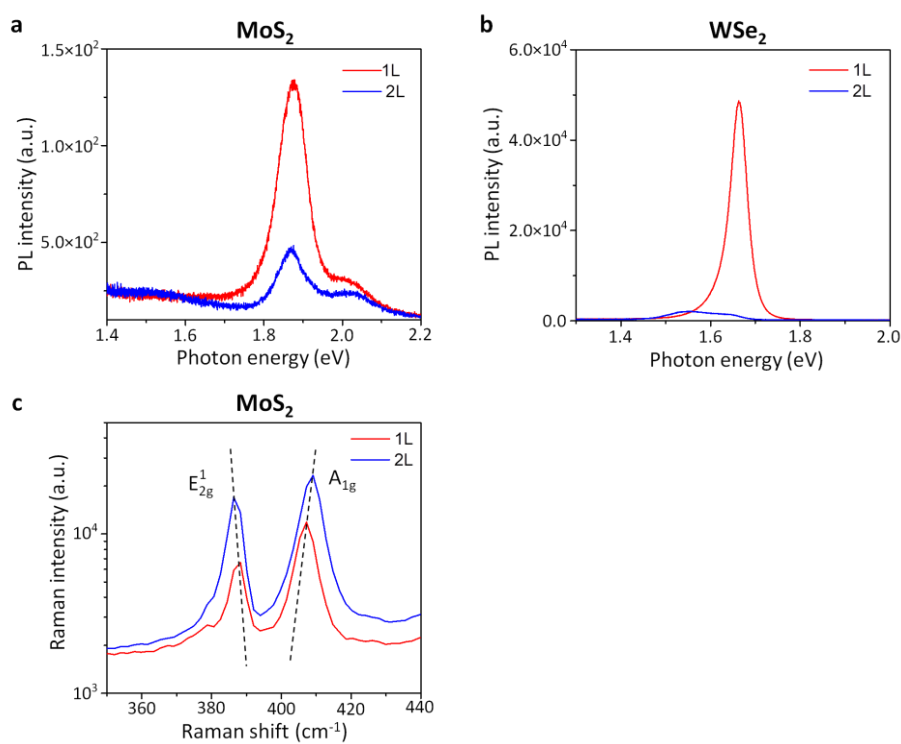

**Figure S6** | PL spectra of **a**, $MoS_2$ and **b**, $WSe_2$ with monolayer and bilayer thicknesses. **c**, Raman spectra of monolayer and bilayer $MoS_2$.

## S7. Spatially resolved photoresponses of graphene-sandwiched *p-n* junction devices with different thicknesses

Figure S7 shows optical images and corresponding photocurrent maps for graphene-sandwiched bilayer and multilayer junction devices. Photocurrent is uniformly observed in the entire junction region where two active TMDC layers overlap with top and bottom graphene contact layers.

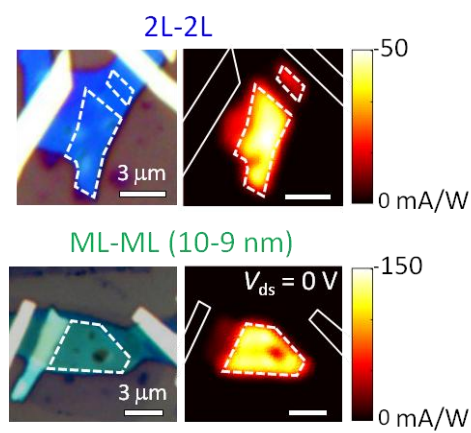

**Figure S7** | Optical images and corresponding photocurrent maps (at $V_{ds}$ = 0 V) of the graphene-sandwiched *p-n* junctions consisting of bilayer-bilayer (2L-2L) and multilayer-multilayer (ML-ML (10-9 nm)). The junction areas and metal electrodes are indicated by the dashed and solid lines, respectively.

## S8. Gate-tuneable photocurrent of graphene-sandwiched *p-n* junction devices with different thicknesses

Figure S8 shows the photoresponse characteristics of graphene-sandwiched p-n junction devices as a function of the gate voltages. Although the monolayer junction ((1L-1L) device exhibits the similar peak behaviour with the lateral device (shown in Fig. 2g), the photoresponsivity of the bilayer (2L-2L) and multilayer (ML-ML) devices vary monotonically with changing the gate voltage. The increases in photoresponsivity with sweeping the gate voltage in the negative direction can be explained by lowering the potential barriers and increasing the band slopes in these heterojunctions[16,17].

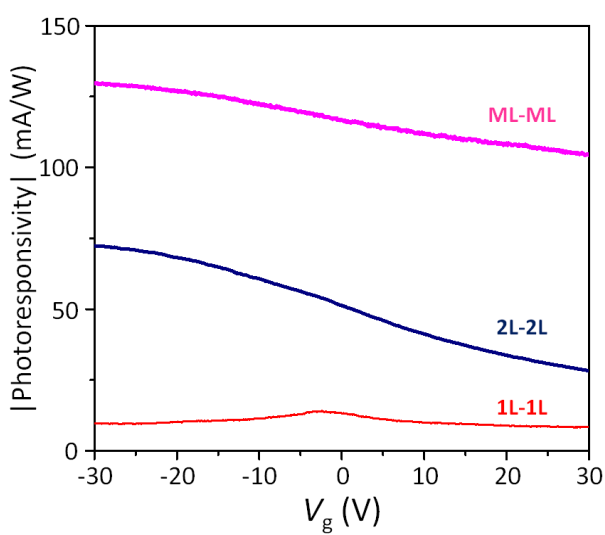

**Figure S8** | Gate-voltage-dependent photoresponse characteristics of graphene-sandwiched *p-n* junctions with different thicknesses. Results are shown for devices (in Fig. 3) composed of monolayer-monolayer (1L-1L), bilayer-bilayer (2L-2L) and multilayer-multilayer (ML-ML

(10-9 nm)) junctions. The measurements are performed under 532-nm laser excitation at $V_{ds} = 0$ V.

## S9. Absorption spectra of the MoS$_2$ / WSe$_2$ heterostructure

Figure S9a shows the absorption spectra of monolayer MoS$_2$ and WSe2 and the MoS$_2$ / WSe$_2$ stack prepared on a transparent fused silica substrate. The absorption can be determined by measuring the reflectivity $R$ and transmissivity $T$ of the heterostructure, and it can thus be deduced by $A = 1 - R - T$. A broadband light emitted from a tungsten halogen lamp is focused onto our micro-sized sample using a Nikon Eclipse TE-2000 optical microscope. The reflected or transmitted light is then dispersed by a grating and measured using a liquid-nitrogen-cooled CCD detector. Reflectivity is determined by measuring the reflection spectra on the sample ($S_{R1}$) and on the substrate ($S_{R2}$) separately, and we obtain $R = S_{R1} / S_{R2} \times (n-1)^2 / (n+1)^2$, where $n$ is the refractive index of the substrate. Transmissivity is determined in a similar fashion by measuring the transmitted spectrum on the sample ($S_{T1}$) and away from the sample ($S_{T1}$), and we obtain $T = S_{T1} / S_{T2} \times 4n / (n+1)^2$. The absorption spectra for monolayer MoS$_2$ and monolayer WSe$_2$ are also shown for comparison. The measured absorption and EQE spectra from the MoS$_2$ / WSe$_2$ heterostructure showed resonance features at the same frequencies, verifying that photocurrent is originated from light absorption in both layers.

Since the absorption spectrum of a layered material strongly depends on the substrate used, we further performed simulation for the MoS$_2$ / WSe$_2$ heterostructures on the 280-nm-thick SiO$_2$ coated Si substrates. Figure S9b shows the simulated absorption spectra from the monolayer and multilayer MoS$_2$ / WSe$_2$ heterostructures. To do that, we extracted the dielectric function of the heterojunction from the measured absorption spectrum for the monolayer heterojunction, and used those values of bulk MoS$_2$ and WSe$_2$ crystals for the multilayer one[18]. By taking into account the substrate interference effect, we then simulate the absorption spectra for two different thicknesses. Both absorption spectra showed the resonance peaks corresponding to excitonic transitions of two materials. However, for the multilayer heterostructure, the peaks became broader than those of the monolayer case because of the reduced lifetime.

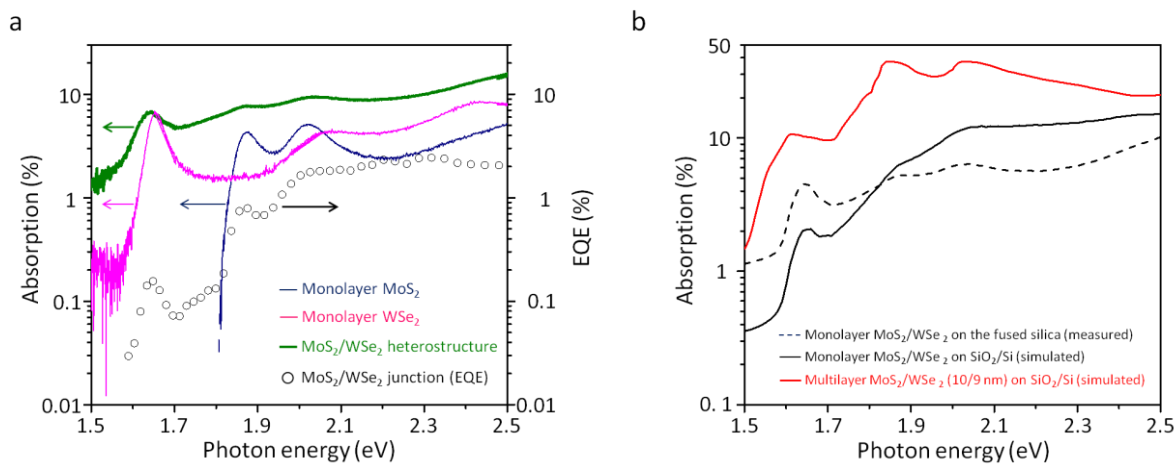

**Figure S9 | a.** Absorption spectra for monolayer $WSe_2$, monolayer $MoS_2$ and the $MoS_2$ / $WSe_2$ heterostructure, measured from the samples prepared on a fused silica substrate. The featured absorption peaks of individual $MoS_2$ and $WSe_2$ monolayers are also observed in the spectrum of the heterostructure. **b.** Simulated absorption spectra for the monolayer $MoS_2$ / $WSe_2$ heterostructure and the multilayer $MoS_2$ / $WSe_2$ (10 / 9 nm) heterostructure on the $SiO_2$ (280 nm) / Si substrate.

## S10. Excitation-power-dependent EQE measurements

Figure S10 shows the measured values of EQE as a function of excitation laser power for graphene-sandwiched *p-n* junction devices. Irrespective of junction thickness, the devices exhibit the linear photoresponse up to the power of ~10 μW, leading to almost constant EQE values. However, the EQE decreases as the excitation power increases above the critical values.

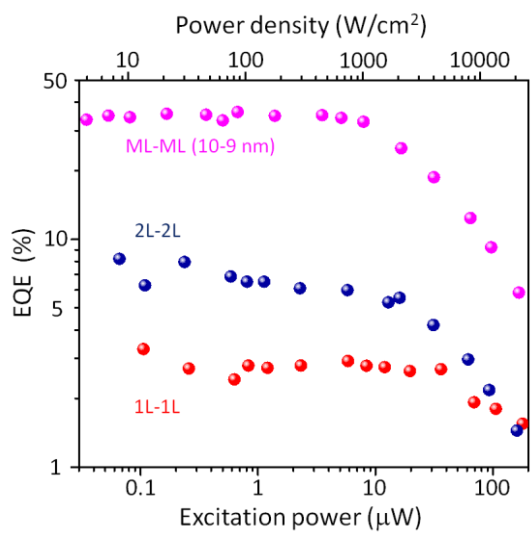

**Figure S10 |** EQE plots as a function of excitation power for graphene-sandwiched *p-n* heterojunctions. Three devices described here include monolayer/monolayer (1L-1L), bilayer/bilayer (2L-2L) and multilayer/multilayer (ML-ML (10-9 nm)) junctions. For the measurements, a 532-nm laser was used as an excitation source.